\documentclass[preprint,showpacs,preprintnumbers,amsmath,amssymb]{revtex4}

\usepackage{graphicx} 
\usepackage{dcolumn}
\usepackage{bm} 

\begin{document}

\title{Contributions of steady heat conduction to the rate of chemical
reaction}

\author{Kim Hyeon-Deuk\footnote{kim@yuragi.jinkan.kyoto-u.ac.jp}, Hisao Hayakawa}

\address{Graduate School of Human and Environmental Studies, Kyoto
 University, Kyoto 606-8501, Japan}

\date{\today}

\begin{abstract}
{We have derived the effect of steady heat flux on the rate of chemical reaction based on the line-of-centers model using the
 explicit velocity distribution function of the steady-state Boltzmann equation for
 hard-sphere molecules to second order. 
It is found that the second-order
 velocity distribution function plays an essential role for the
 calculation of it. 
We have also compared our result with those from the
 steady-state Bhatnagar-Gross-Krook(BGK) equation and information
 theory, and found no qualitative differences among them. 
}
\end{abstract}


\maketitle

\newpage

\section{Introduction}
Chemical reactions in gases has been studied with the aid of gas kinetic 
theory\cite{chapman,resi,leb,present}. 
Present proposed the line-of-centers model for chemical reaction in gases, which has been accepted as a standard model to describe the chemical
reaction in gases.\cite{present,present1,present3,present4,present5,present6,fort2,fort10,fort,fort1,fort3,eu,sk1,sk2,sk3,present2,present10,sigma} 

Under nonequilibrium situations such as gases under the
heat conduction and the shear flow, their nonequilibrium effects on the rate of
chemical reaction have attracted attention among researchers.\cite{fort,fort1,fort3,eu,sk1,sk2,sk3,net,net1}
However, to derive the effect of steady heat flux on the rate of
chemical reaction in the line-of-centers model, we need the explicit velocity distribution function of the steady-state Boltzmann equation for
hard-sphere molecules to second order in density and temperature
 gradient. 
To our knowledge, nobody had derived the explicit velocity
 distribution function of the Boltzmann
equation for hard-sphere molecules to second order. 
Although Burnett determined the second-order pressure tensor for the Boltzmann
equation\cite{chapman}, he had not derived the
explicit second-order velocity distribution function of the Boltzmann
equation.\cite{burnett,kim1}
This is a result of mathematical difficulties, as was indicated by Fort and Cukrowski.\cite{fort,fort1} 
Therefore, none has succeeded to obtain the correct reaction rate of Present's
model except for Fort and Cukrowski who adopted information theory\cite{jou} as
the nonequilibrium velocity distribution function to second order.\cite{fort,fort1} 

Recently, we have derived the explicit velocity distribution function of the steady-state Boltzmann equation for
 hard-sphere molecules to second order.\cite{kim1} 
In this letter, we apply the explicit velocity distribution function
to the calculation of the rate of chemical reaction in the
line-of-centers model under steady heat conduction. 

\section{Simple model for chemical reaction}
In the early stage of a chemical reaction between monatomic molecules: 
\begin{equation}
A + A \rightarrow \mathrm{products}, \label{cr10}
\end{equation}
the rate of chemical reaction is not affected by the existence of
products.\cite{pri} 
From the viewpoint of kinetic collision theory\cite{chapman,resi,leb,present}, the rate of chemical reaction
(\ref{cr10}) can be described as 
\begin{equation}  
R=
\int d{\bf v} \int d{\bf v}_{\mathrm{1}} \int d{\bf \Omega} \int f f_{\mathrm{1}} g \sigma(g),\label{cr20}
\end{equation}
where ${\bf v}$ and ${\bf v}_{\mathrm{1}}$ are the velocities of the
molecules, $g=|{\bf v}-{\bf v}_{\mathrm{1}}|$ their relative speed, ${\bf \Omega}$ the solid angle, 
$f=f({\bf r},{\bf v})$ and $f_{\mathrm{1}}=f({\bf r},{\bf
v}_{\mathrm{1}})$ are the distributions of ${\bf v}$ and ${\bf
v}_{\mathrm{1}}$ at ${\bf r}$, respectively. 

We adopt the line-of-centers model as the differential cross-section of
chemical reaction $\sigma(g)$.\cite{present,present1,present3,present4,present5,present6,fort2,fort10,fort,fort1,fort3,eu,sk1,sk2,sk3,present2,present10,sigma}  
The model contains
\begin{eqnarray}  
\sigma(g)= \left\{
\begin{array}{@{\,}ll}
0 & \quad g < \sqrt{\frac{4 E^{*}}{m}} \\
\frac{d^{2}}{4}\left(1-\frac{4 E^{*}}{m g^{2}}\right) & \quad g \ge \sqrt{\frac{4 E^{*}}{m}} \\
\end{array}
\right.,
\label{cr30}
\end{eqnarray} 
with $m$ mass of the molecules and $E^{*}$ the
threshold energy of the chemical reaction. 
$d$ is regarded as a distance between centers of monatomic
molecules at contact.\cite{present,sigma}

In order to calculate the rate of chemical reaction (\ref{cr20}), 
we expand the velocity distribution function $f$ to second order as 
\begin{equation}  
f=f^{(0)}+f^{(1)}+f^{(2)}=f^{(0)}(1+\phi^{(1)}+\phi^{(2)}), \label{cr40}
\end{equation} 
 around the local Maxwellian, $f^{(0)}=n(m/2\pi\kappa
T)^{3/2}\exp[-m {\bf v}^{2}/2 \kappa T]$, with $n$ the density of molecules, $\kappa$ the
Boltzmann constant and $T$ the temperature defined from the kinetic
energy. 
Substitution of eq.(\ref{cr40}) into eq.(\ref{cr20}) leads to 
\begin{equation}  
R=R^{(0)}+R^{(1)}+R^{(2)}, \label{cr50}
\end{equation}
up to second order. 
The zeroth-order term of $R$,
\begin{equation}  
R^{(0)}=\int d{\bf v} \int d{\bf v}_{\mathrm{1}} \int d{\bf \Omega} \int f^{(0)} f_{\mathrm{1}}^{(0)} g \sigma(g)=4n^{2}\sigma^{2}\left(\frac{\pi \kappa T}{m}\right)^{\frac{1}{2}}
e^{-\frac{E^{*}}{\kappa T}} , \label{cr60}
\end{equation}
corresponds to the rate of chemical reaction of the equilibrium theory. 
Similarly, the first-order term of $R$ is obtained as 
\begin{equation}  
R^{(1)}=\int d{\bf v} \int d{\bf v}_{\mathrm{1}} \int d{\bf \Omega} \int f^{(0)} f_{\mathrm{1}}^{(0)}[\phi^{(1)}+\phi^{(1)}_{\mathrm{1}}] g \sigma(g), \label{cr70}
\end{equation}
where $R^{(1)}$ does not appear because $\phi^{(1)}$ is an odd functions
of ${\bf c}$, as will be shown in the next section.  
The second-order term of $R$, \textit{i.e.} $R^{(2)}$, is divided into 
\begin{equation}  
R^{(2,A)}=\int d{\bf v} \int d{\bf v}_{\mathrm{1}} \int d{\bf \Omega} \int f^{(0)} f_{\mathrm{1}}^{(0)}\phi^{(1)} \phi^{(1)}_{\mathrm{1}} g \sigma(g), \label{cr80}
\end{equation}
and
\begin{equation}  
R^{(2,B)}=\int d{\bf v} \int d{\bf v}_{\mathrm{1}} \int d{\bf \Omega} \int f^{(0)} f_{\mathrm{1}}^{(0)}[\phi^{(2)}+\phi^{(2)}_{\mathrm{1}}] g \sigma(g). \label{cr90}
\end{equation}
Since the integrations (\ref{cr80}) and (\ref{cr90}) have the cutoff
from eq.(\ref{cr30}), the explicit forms of $\phi^{(1)}$ and $\phi^{(2)}$ of the steady-state Boltzmann equation for
hard-sphere molecules are required to calculate $R^{(2,A)}$ and
$R^{(2,B)}$, respectively. 

\section{Explicit form of the velocity distribution function to second order}
We introduce the explicit form of the velocity distribution function of the steady-state Boltzmann equation for
 hard-sphere molecules to second
order in the density and the temperature gradient along
 $x$-axis\cite{kim1}:  
\begin{eqnarray}  
\phi^{(1)}=-\frac{4J_{x}}{5 b_{11} n\kappa T}\left(\frac{m}{2\kappa T}\right)^{\frac{1}{2}}\sum_{r\ge 1}r! b_{1r}c_{x}\Gamma(r+\frac{5}{2}) S^{r}_{\frac{3}{2}}({\bf c}^{2})
, \label{cr100}
\end{eqnarray}
which is an odd function of ${\bf c}$, and 
\begin{eqnarray}  
\phi^{(2)}=\frac{4096mJ_{x}^{2}}{5625b_{11}^{2}n^{2}\kappa^{3}T^{3}}\{\sum_{r\ge 2}r! b_{0r}\Gamma(r+\frac{3}{2})S^{r}_{\frac{1}{2}}({\bf c}^{2})+ \nonumber \\
\sum_{r\ge 0}r! b_{2r}(2c_{x}^{2}-c_{y}^{2}-c_{z}^{2})\Gamma(r+\frac{7}{2})S^{r}_{\frac{5}{2}}({\bf c}^{2})\}
, 
\label{cr110}
\end{eqnarray}
with the scaled velocity ${\bf c}\equiv (m/2\kappa T)^{1/2}{\bf v}$. 
Here $\Gamma(X)$ represents the Gamma function and $S_{k}^{r}(X)$ is a Sonine
polynomial, the definition and the orthogonality properties of which are written in our paper.\cite{kim1}
 $J_{x}=-75 b_{11}\kappa^{3/2}T^{1/2}\partial_{x}
 T/64\pi^{1/2}m^{1/2}d^{2}$ denotes the steady heat
 flux with $\partial_{x} \equiv\partial /\partial x$. 
The numerical values for $b_{1r}$, $b_{0r}$ and $b_{2r}$ are listed in
Table \ref{bkr}. 
Note that, though we have derived them in forms of fractions, we use
them in forms of four significant figures, since the forms of the fractions are
too complicated. 
We have derived each of $\phi^{(1)}$ and $\phi^{(2)}$ to $7$th
approximation of Sonine polynomials, though the complete forms of
$\phi^{(1)}$ and $\phi^{(2)}$ for hard-sphere molecules are
the sums of an infinite series of Sonine polynomials. 
However, we have confirmed that $\phi^{(2)}$ does not converge to $4$th Sonine
approximation but both of $\phi^{(1)}$ and $\phi^{(2)}$ almost converge to $7$th Sonine
approximation\cite{kim1}, so that, in this letter, we will show the
results calculated from $\phi^{(1)}$ and $\phi^{(2)}$ for $7$th approximation of Sonine polynomials. 
We note that Burnett had determined only $b_{1r}$ and $b_{2r}$ to $4$th Sonine
approximation: $b_{2r}$ is related to
the second-order pressure tensor.\cite{burnett} 
Therefore, we could not calculate the effect of steady heat flux on the rate of
chemical reaction with the line-of-centers model if we did not derive
the explicit forms of $\phi^{(1)}$ and $\phi^{(2)}$, especially $b_{0r}$ in ref.\cite{kim1}. 

In order to compare the results from the steady-state Boltzmann equation with those from the
 steady-state Bhatnagar-Gross-Krook(BGK) equation and information theory, we also use the 
 explicit forms of $\phi^{(1)}$ and $\phi^{(2)}$ for them. 
The expressions of $\phi^{(1)}$ and $\phi^{(2)}$ for the steady-state
BGK equation can be reduced from the general form of the Chapman-Enskog solution for the steady-state BGK equation to arbitrary order\cite{santos,santos1}, while those for information theory\cite{jou} were used by Fort and Cukrowski.\cite{fort,fort1}
We mention that $\phi^{(1)}$ for the steady-state BGK equation is an odd
 function of ${\bf c}$ and identical with that for information theory,
 while $\phi^{(2)}$ are different from each other. 

\section{Results: rate of chemical reaction}\label{results}
Inserting $\phi^{(1)}$ and $\phi^{(2)}$ of
eqs.(\ref{cr100}) and (\ref{cr110}) and the corresponding velocity
distribution functions for the steady-state BGK equation and
information theory into eqs.(\ref{cr80}) and (\ref{cr90}), and
performing the integrations with the chemical reaction cross-section
(\ref{cr30}), we finally obtain the nonequilibrium effects on
the rate of chemical reaction based on the line-of-centers model. 
The expressions of $R^{(2,A)}$ and $R^{(2,B)}$ become
\begin{eqnarray}  
R^{(2,A)}=\frac{\sigma^{2} m J_{x}^{2}}{\kappa^{3}T^{3}}\left(\frac{\pi \kappa T}{m}\right)^{\frac{1}{2}}
e^{-\frac{E^{*}}{\kappa T}} \{\sum_{r\ge 0} \alpha_{r} \left(\frac{E^{*}}{\kappa T}\right)^{r}\}
, 
\label{cr120}
\end{eqnarray}
and
\begin{eqnarray}  
R^{(2,B)}=\frac{\sigma^{2} m J_{x}^{2}}{\kappa^{3}T^{3}}\left(\frac{\pi \kappa T}{m}\right)^{\frac{1}{2}}
e^{-\frac{E^{*}}{\kappa T}} \{\sum_{r\ge 0} \beta_{r} \left(\frac{E^{*}}{\kappa T}\right)^{r}\}
, 
\label{cr130}
\end{eqnarray}
respectively. 
The numerical values for $\alpha_{r}$ and $\beta_{r}$ are listed in
Tables \ref{alpha} and \ref{beta}, respectively. 
We have found that $R^{(2,B)}$ for the steady-state Boltzmann equation is determined only by the terms of
$b_{0r}$ in $\phi^{(2)}$ of eq.(\ref{cr110}) which Burnett\cite{burnett} had not
derived: the terms of $b_{0r}$ we have derived in ref.\cite{kim1} are indispensable for 
the calculation of $R^{(2,B)}$. 

The explicit expressions in eqs.(\ref{cr120}) and (\ref{cr130}) for
information theory were already given by Fort and Cukrowski.\cite{fort,fort1}  
Though they were
interested in forms of $Q_{A}\equiv R^{(2,A)}/R^{(0)}$ and
$Q_{B}\equiv R^{(2,B)}/R^{(0)}$\cite{fort,fort1}, in the present letter,
we focus on the nonequilibrium effects on
the rate of chemical reaction in the forms of $R^{(2,A)}$ and
$R^{(2,B)}$. 
This is because the forms of $Q_{A}$ and $Q_{B}$ for the steady-state
Boltzmann equation do not converge even when we adopt any higher order
approximation of Sonine polynomials. 
The upper limit of $r$ in
eqs.(\ref{cr120}) and (\ref{cr130}) is directly related to the order of
the approximation of Sonine polynomials in eqs.(\ref{cr100}) and
(\ref{cr110}): $S_{k}^{r}({\bf c}^{2})$ includes the term
of ${\bf c}^{2r}$. 
We emphasize that, however, the values of $R^{(2,A)}$ and $R^{(2,B)}$ for the
steady-state Boltzmann equation converge to $7$th approximation of
Sonine polynomials. 
It should be also mentioned that there appear artificial qualitative
differences in both $Q_{A}$ and $Q_{B}$ for the steady-state Boltzmann equation,
the steady-state BGK equation and information theory for large $E^{*}/\kappa T$. 

\begin{figure}[hp]
\includegraphics[width=15.cm]{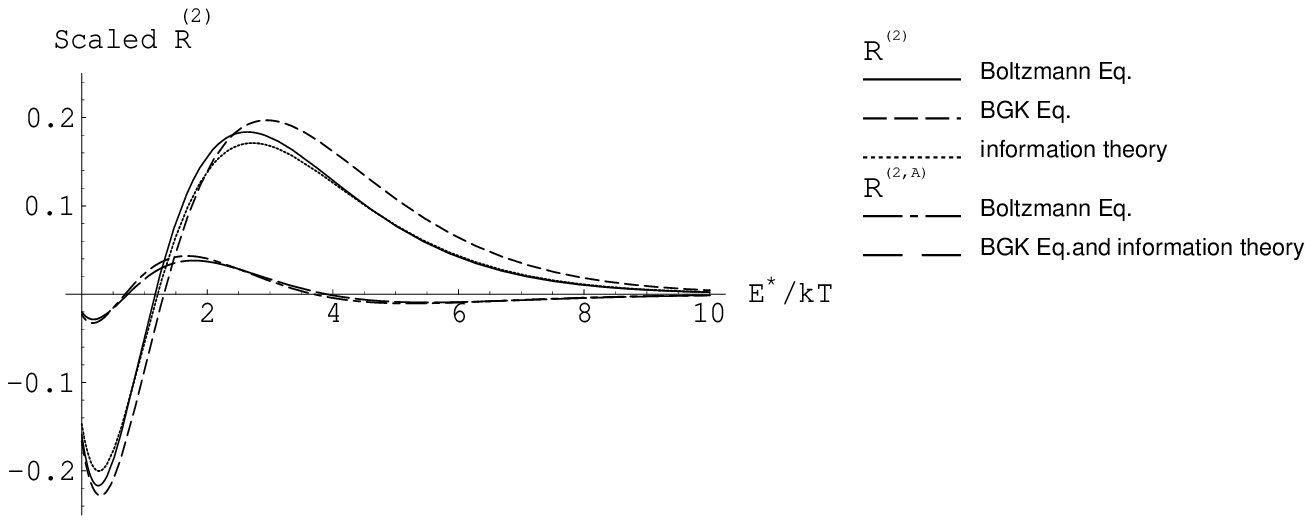} 
\caption{Scaled $R^{(2)}$ compared to scaled $R^{(2,A)}$ as a function
 of $E^{*}/\kappa T$ for the line-of-centers model. } 
\label{CR}
\end{figure}

The graphical results of $R^{(2)}$ compared with those of $R^{(2,A)}$
are provided in
Fig.\ref{CR}. 
Both of $R^{(2)}$ and $R^{(2,A)}$ in Fig.\ref{CR} are scaled by $\pi^{1/2}d^{2} m^{1/2}
J_{x}^{2}/\kappa^{5/2}T^{5/2}$. 
Note that $R^{(2)}$ is the sum of $R^{(2,A)}$ and $R^{(2,B)}$ in eqs.(\ref{cr120}) and
(\ref{cr130}). 
As Fig.\ref{CR} shows, it is clear that $R^{(2,B)}$ plays an essential role for the evaluation
of $R^{(2)}$. 
We have also found that there are no qualitative differences in $R^{(2)}$ of the steady-state Boltzmann equation,
the steady-state BGK equation and information theory, while $R^{(2,A)}$
also exhibits a slight deviation from each other. 
This deviation in $R^{(2,A)}$ would not be observed if we adopted
$\phi^{(1)}$ of the steady-state Boltzmann equation for 
the lowest approximation of Sonine polynomials as in the previous papers\cite{fort3,eu}. 
This is because $\phi^{(1)}$ of the steady-state Boltzmann equation for 
the lowest Sonine approximation is identical with the precise $\phi^{(1)}$ of the steady-state BGK
 equation and information theory. 
It should be mentioned that we have found qualitative differences among
these theories in some thermodynamic quantities, \textit{e.g.} the second-order pressure tensor.\cite{kim1,kim2} 
The nonequilibrium effects on the rate of chemical reaction is an
insensitive quantity to the differences among the three theories. 

\section{Discussion}\label{discussion}
 $\phi^{(2)}$ is indispensable for the calculation of the nonequilibrium
 effects on the rate of chemical reaction, since $R^{(1)}$
 does not appear and $R^{(2,B)}$ is remarkably
 larger than $R^{(2,A)}$ as Fig.\ref{CR} shows.  
Thus, the nonequilibrium effect on the rate of chemical reaction
will substantiate significance of the second-order
 coefficients in the solution of the steady-state Boltzmann equation,
 although their importance has been demonstrated only for
descriptions of shock wave profiles and sound
propagation phenomena\cite{foch,shocksound,shocksound1}.  
We should emphasize that the calculation of the second-order pressure
 tensor and the application of it to shock wave profiles and sound
propagation do not require the explicit form of the second-order
 velocity distribution function: the second-order coefficients derived
 by Burnett do not include any terms of $b_{0r}$ in $\phi^{(2)}$ of eq.(\ref{cr110}).\cite{burnett,kim1} 
The nonequilibrium effects we consider appear mainly
in the ranges of small $E^{*}/\kappa T$ where the chemical reaction often
 occurs, so that our results should be compared with experimental results 
in the early stage of the chemical reaction\cite{pri}; 
a large additional effect due to some modification of the velocity
distribution function because of the chemical reaction itself becomes
significant as the chemical reaction proceeds\cite{sk2,sk3}. 

\begin{figure}[hp]
\includegraphics[width=10.cm]{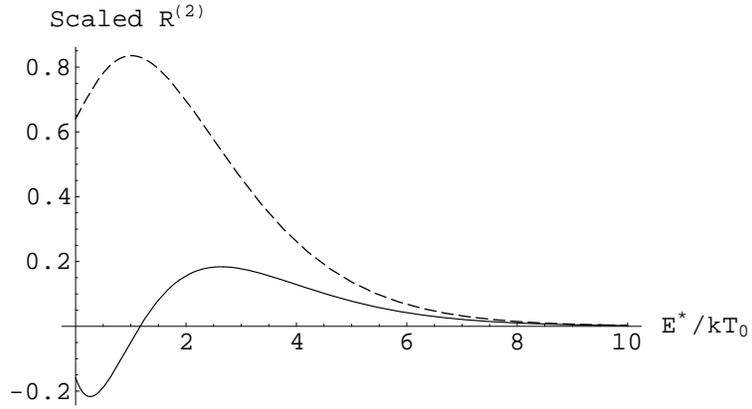} 
\caption{Comparison of scaled $R^{(2)}$ from the steady-state Boltzmann
 equation with $T=T_{0}$ (the solid line) and that with $\theta=T_{0}$ (the dashed line) as
 a function of $E^{*}/\kappa T_{0}$. Note that the former is identical with
 the result shown by the solid line in Fig.1 with $T=T_{0}$. }
\label{Theta}
\end{figure}

We also propose a \textit{thermometer} of a monatomic
dilute gas system under steady heat flux. 
We mean that we can measure the temperature $T$ around a heat bath at
$T_{0}$ in the nonequilibrium
steady-state system indirectly with the aid of the
nonequilibrium effect on the rate of chemical reaction as follows. 
The nonequilibrium effect on the rate of chemical
reaction in the early stage around the heat bath can be measured. 
Thus, one can compare the experimental result with the theoretical
result shown in
 Fig.\ref{CR} with $T=T_{0}$. 
The difference between the former and the latter will indicate that the
temperature $T$ around the heat bath is not
identical with $T_{0}$, that is, $T=T_{0}+\Delta$. 
Here $\Delta$ should depend upon the heat flux in general. 
Substituting this temperature expression into the explicit expressions
of $R^{(0)}$ in eq.(\ref{cr60}), we obtain a new correction term
\begin{eqnarray}  
R^{(\mathrm{new})}=\frac{2 n^{2} \sigma^{2} \Delta}{T_{0}}\left(\frac{\pi \kappa T_{0}}{m}\right)^{\frac{1}{2}}\left(1+\frac{2E^{*}}{\kappa T_{0}}\right)e^{-\frac{E^{*}}{\kappa T_{0}}} 
, 
\label{cr140}
\end{eqnarray}
as the nonequilibrium effect on the rate of
chemical reaction besides $R^{(2)}$ in Fig.\ref{CR} with $T=T_{0}$. 
At last, we can estimate the gap $\Delta$ so as to make the new
correction term match the experimental result. 
For example, if $\Delta$ in eq.(\ref{cr140}) is proportional to the heat 
flux, we will confirm the 
relevancy of the slip effect\cite{sone,cercignani}. 
If $\Delta$ in eq.(\ref{cr140}) is identical with $2 m
J_{x}^{2}/5n^{2}\kappa^{3}T_{0}^{2}$, \textit{i.e.} $1/T=1/T_{0}-2 m
J_{x}^{2}/5n^{2}\kappa^{3}T_{0}^{4}$,  
the experimental result will agree with the theoretical
result from the steady-state Boltzmann equation with the
\textit{nonequilibrium temperature} $\theta=T_{0}$ predicted by Jou \textit{et
al.}\cite{jou}. 
We show the comparison of the theoretical results, \textit{i.e.} $R^{(2)}$ for the
steady-state Boltzmann equation with
$T=T_{0}$ and that with $\theta=T_{0}$, in Fig.\ref{Theta}, where $R^{(2)}$ is scaled by $\pi^{1/2}d^{2} m^{1/2}
J_{x}^{2}/\kappa^{5/2}T^{5/2}_{0}$. 
We note that one can obtain the explicit form of $R^{(2)}$ with $\theta=T_{0}$ expressed as the dashed line
in Fig.\ref{Theta} as the sum of $R^{(2)}$ from eqs.(\ref{cr120}) and
(\ref{cr130}) for the steady-state Boltzmann equation with
$T=T_{0}$ and $R^{(\mathrm{new})}$ from eq.(\ref{cr140}) with $\Delta=2 m
J_{x}^{2}/5n^{2}\kappa^{3}T_{0}^{2}$.  
We have found that there is a significant difference between 
$R^{(2)}$ with $T=T_{0}$ and that with $\theta=T_{0}$ for small $E^{*}/\kappa T_{0}$. 
This significant difference also appears in $R^{(2)}$ for the steady-state BGK
equation and information theory. 
Similar difference in $R^{(2)}$ for
information theory would also follow from the results of Fort and Cukrowski\cite{fort1},
although they were not interested in the ranges of small
$E^{*}/\kappa T$. 
We emphasize that, however, our proposal written in this paragraph differs from that by
Fort and Cukrowski\cite{fort1} which uses a given value measured
directly by a thermometer in order to determine $T$ or $\theta$ regardless of the temperature of the heat bath. 

\section*{acknowledgments}
This study has been supported partially by the Hosokawa Powder Technology
Foundation and the Inamori Foundation.

\newpage

\begin{table}[hp]
\begin{ruledtabular}
\caption{\label{bkr}Numerical constants $b_{1r}$, $b_{0r}$ and $b_{2r}$ in
 eqs.(\ref{cr100}) and (\ref{cr110}). }
\begin{tabular}{|c|c|c|c|} \hline
{$r$}&{$b_{1r}$}&{$b_{0r}$}&{$b_{2r}$} \\ \hline
{$0$}&{$0$}&{$1$}&{$-3.320\times 10^{-2}$} \\ \hline
{$1$}&{$1.025$}&{$0$}&{$-1.276\times 10^{-1}$} \\ \hline
{$2$}&{$4.892\times 10^{-2}$}&{$4.380\times 10^{-1}$}&{$6.414\times 10^{-2}$} \\ \hline
{$3$}&{$3.715\times 10^{-3}$}&{$-5.429\times 10^{-2}$}&{$5.521\times 10^{-3}$} \\ \hline
{$4$}&{$2.922\times 10^{-4}$}&{$-4.098\times 10^{-3}$}&{$4.214\times 10^{-4}$} \\ \hline
{$5$}&{$2.187\times 10^{-5}$}&{$-3.184\times 10^{-4}$}&{$3.106\times 10^{-5}$} \\ \hline
{$6$}&{$1.492\times 10^{-6}$}&{$-2.087\times 10^{-5}$}&{$1.861\times 10^{-6}$} \\ \hline
{$7$}&{$8.322\times 10^{-8}$}&{$-$}&{$-$} \\ \hline
\end{tabular}
\end{ruledtabular}
\end{table}

\begin{table}[hp]
\begin{ruledtabular}
\caption{\label{alpha}Numerical constants $\alpha_{r}$ in
 eq.(\ref{cr120}). }
\begin{tabular}{|c|c|c|} \hline
{$r$}&{Boltzmann Eq.}&{BGK Eq. and Information Theory} \\ \hline
{$0$}&{$-2.292\times 10^{-2}$}&{$-\frac{1}{50}$} \\ \hline
{$1$}&{$-1.448\times 10^{-1}$}&{$-\frac{3}{25}$} \\ \hline
{$2$}&{$3.223\times 10^{-1}$}&{$\frac{6}{25}$} \\ \hline
{$3$}&{$-9.834\times 10^{-2}$}&{$-\frac{4}{75}$} \\ \hline
{$4$}&{$7.919\times 10^{-3}$}&{$-$} \\ \hline
{$5$}&{$-6.752\times 10^{-4}$}&{$-$} \\ \hline
{$6$}&{$5.298\times 10^{-5}$}&{$-$} \\ \hline
{$7$}&{$-3.584\times 10^{-6}$}&{$-$} \\ \hline
{$8$}&{$2.039\times 10^{-7}$}&{$-$} \\ \hline
{$9$}&{$-9.599\times 10^{-9}$}&{$-$} \\ \hline
{$10$}&{$3.613\times 10^{-10}$}&{$-$} \\ \hline
{$11$}&{$-1.032\times 10^{-11}$}&{$-$} \\ \hline
{$12$}&{$2.110\times 10^{-13}$}&{$-$} \\ \hline
{$13$}&{$-2.887\times 10^{-15}$}&{$-$} \\ \hline
{$14$}&{$2.352\times 10^{-17}$}&{$-$} \\ \hline
{$15$}&{$-8.578\times 10^{-20}$}&{$-$} \\ \hline
\end{tabular}
\end{ruledtabular}
\end{table}

\begin{table}[hp]
\begin{ruledtabular}
\caption{\label{beta}Numerical constants $\beta_{r}$ in eq.(\ref{cr130}).}
\begin{tabular}{|c|c|c|c|} \hline
{$r$}&{Boltzmann Eq.}&{BGK Eq.}&{Information Theory} \\ \hline
{$0$}&{$-1.361\times 10^{-1}$}&{$-\frac{11}{75}$}&{$-\frac{19}{150}$} \\ \hline
{$1$}&{$-5.094\times 10^{-1}$}&{$-\frac{38}{75}$}&{$-\frac{7}{15}$} \\ \hline
{$2$}&{$3.968\times 10^{-1}$}&{$\frac{4}{15}$}&{$\frac{26}{75}$} \\ \hline
{$3$}&{$5.805\times 10^{-2}$}&{$\frac{8}{75}$}&{$\frac{4}{75}$} \\ \hline
{$4$}&{$-2.811\times 10^{-3}$}&{$-$}&{$-$} \\ \hline
{$5$}&{$1.039\times 10^{-4}$}&{$-$}&{$-$} \\ \hline
{$6$}&{$-1.808\times 10^{-6}$}&{$-$}&{$-$} \\ \hline
\end{tabular}
\end{ruledtabular}
\end{table}

\end{document}